# The Polstar High Resolution Spectropolarimetry MIDEX Mission

Paul A. Scowen*[a], Ken Gayley[b], Coralie Neiner[c], Gopal Vasudevan[d], Robert Woodruff[e], Richard Ignace[f], Roberto Casini[g], Tony Hull[h], Alison Nordt[d], H. Philip Stahl[i]

[a]NASA Goddard Space Flight Center, 8800 Greenbelt Rd., Greenbelt, MD 20771; [b]Dept. of Physics & Astronomy, University of Iowa, Iowa City, Iowa 52242-1479; [c]Observatoire de Meudon, 92195 Meudon Cedex, France; [d]Lockheed Martin Advanced Technology Center, 3251 Hanover Street, Palo Alto, CA 94034; [e]Woodruff Consulting, 2081 Evergreen Ave, Boulder, CO 80304; [f]Dept of Physics & Astronomy, East Tennessee State University, 1276 Gilbreath Dr., Box 70300, Johnson City, TN 37614-1700; [g]High Altitude Observatory, NCAR, 3090 Center Green Drive, Boulder, CO 80301; [h]Dept of Physics & Astronomy, University of New Mexico, 210 Yale Blvd NE, Albuquerque, NM 87106; [i]NASA Marshall Space Flight Center, Mail Stop ES-34, Huntsville, AL 35812

## ABSTRACT

The Polstar mission will provide for a space-borne 60cm telescope operating at UV wavelengths with spectropolarimetric capability capturing all four Stokes parameters (intensity, two linear polarization components, and circular polarization). Polstar's capabilities are designed to meet its goal of determining how circumstellar gas flows alter massive stars' evolution, and finding the consequences for the stellar remnant population and the stirring and enrichment of the interstellar medium, by addressing four key science objectives. In addition, Polstar will determine drivers for the alignment of the smallest interstellar grains, and probe the dust, magnetic fields, and environments in the hot diffuse interstellar medium, including for the first time a direct measurement of the polarized and energized properties of intergalactic dust. Polstar will also characterize processes that lead to the assembly of exoplanetary systems and that affect exoplanetary atmospheres and habitability. Science driven design requirements include: access to ultraviolet bands: where hot massive stars are brightest and circumstellar opacity is highest; high spectral resolution: accessing diagnostics of circumstellar gas flows and stellar composition in the far-UV at 122-200nm, including the NV, SiIV, and CIV resonance doublets and other transitions such as NIV, AlIII, HeII, and CIII; polarimetry: accessing diagnostics of circumstellar magnetic field shape and strength when combined with high FUV spectral resolution and diagnostics of stellar rotation and distribution of circumstellar gas when combined with low near-UV spectral resolution; sufficient signal-to-noise ratios: ~$10^3$ for spectropolarimetric precisions of 0.1% per exposure; ~$10^2$ for detailed spectroscopic studies; ~10 for exploring dimmer sources; and cadence: ranging from 1-10 minutes for most wind variability studies, to hours for sampling rotational phase, to days or weeks for sampling orbital phase. The ISM and exoplanet science program will be enabled by these capabilities driven by the massive star science.

## 1. INTRODUCTION

Access to the far ultraviolet (FUV) spectroscopically provides access to an entire suite of resonant diagnostic lines that can map a large variety of physical phenomena from gas and plasma electron densities and temperatures, to gas mass flows, stellar rotational speeds, plasma optical depths, and metallicities of material in the gas phase [1]. Add to this high resolution spectroscopy and analysis of magnetic field strengths and topologies can be added to the mix. Measurements of these diagnostics can be done either in emission or in absorption depending the physical nature of the system under scrutiny. These merits have been long understood and have made UV spectroscopy of great value to the astronomical community through missions and instruments such as IUE [2], HUT [3], EUVE [4], HST-STIS [5], FUSE [6], and HST-COS [7]. In Polstar we are proposing the addition of high precision (0.1%) polarimetric measurement of all 4 Stokes parameters to this already-impressive suite of diagnostic tools to deliver a mission that will detail for the first time the role of environment and magnetic fields on the evolution of massive stars, reveal processes that energize the interstellar medium, and provide insight into the processes that govern the formation and evolution of exoplanetary systems. The legacy of Polstar extends to earlier missions such as WUPPE [8] and HST-FOS [9] which provided pathfinding capabilities in polarimetry, but with recent advances in polarimetric techniques [10], UV coatings [11] and UV solid state detectors [12], married with innovative optical design, we stand ready to deliver orders of magnitude improvement in performance over those earlier missions, some 25 years after they were in service. The time is right for a renaissance in FUV spectropolarimetry and this paper outlines both the design features that enable these advances and the science we initially seek to advance, that places requirements on the design, and takes advantage of the projected performance of the observatory we will propose.

*paul.a.scowen@nasa.gov; phone 1 602 617-3330; science.gsfc.nasa.gov/astrophysics/exoplanets/

## 2. INSTRUMENT DESIGN

### 2.1 Instrument Executive Summary

Polstar science requirements drive spectroscopy of individual and binary stars and polarimetric capability with a passband in the vacuum ultraviolet (122 nm to 320 nm). These requirements are met with a point source spectrograph that incorporates a modulator and analyzer in the design.

The Polstar instrument (Fig. 1) is a classic two mirror all reflective Cassegrain telescope with diffraction limited performance at 1200 nm. However, spectropolarimetric requirements down to 122 nm require maximizing throughput of photons. Special coatings and minimum reflections are mandatory, and uncommon attention to cleanliness throughout the entire AI&T cycle, and even providing decontamination heaters upon commissioning and periodically through lifetime in space.

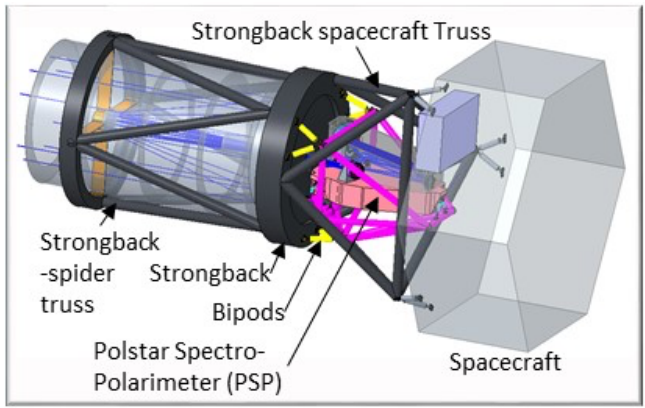

Figure 1 - Designed to cost Polstar instrument delivers the STM performance

Modulator waveplates optimized for the polarimetric waveband (122nm to ~320 nm) in a rotary mechanism (IRIS mechanism heritage) and analyzer is placed in line with the output of the telescope near its Cassegrain focus. The spectropolarimeter instrument is sensitive to wavelengths between 115 nm to > 320 nm. The instrument comprises of two channels. The high-resolution spectrograph is a cross-dispersed echelle grating with a resolving power of 30,000 with Nyquist sampled resolution elements. The detector is a δ-doped standard 4k x 4k Teledyne e2v CCD 272-84 with sufficient active area for the needs of Polstar.

A mirrored thin plate with a 30 micron (corresponding to 0.793 arcsec field) pinhole is placed just ahead of the modulator (Fig. 2) near the Cassegrain focus. A guide star within 3 arcmins from the target object is imaged onto a CMOS fine guidance camera (FGS). The fine guidance camera is used for guiding to maintain instrument pointing. It is also used as a slit-jaw imager and is used for relative photometry of the target with offset pointing (~ 2 arcsec) of the telescope.

The spectropolarimeter is warm-biased and temperature-controlled to operate at room temperature ≤±5°C, depending on the results of a detailed error budget and Structural Thermal Optical Performance (STOP) analysis.

The optical telescope assembly (OTA) is an enhanced $MgF_2$/Aluminum coated two mirror classic Cassegrain telescope with a Secondary mirror on a tip/tilt and defocus mechanism based on IRIS/SUVI and flight DoD programs. The Polstar Spectro-Polarimeter aft optics (PSP) is on a thermally controlled optical bench to maintain dimensional stability between the pinhole and its conjugate on the science camera. Careful monitoring of the mass properties is performed to minimize pointing jitter and drift during the observation to within the error tolerances for the instrument.

A thermal shield thermally isolates the OTA and PSP from surrounding space and serves as a structural connection between the door assembly and spacecraft. It also provides a mounting surface for the two radiators. For high strength and low weight, the thermal shield is constructed from aluminum inner and outer facesheets with an aluminum honeycomb core.

The number of mechanisms used are kept to a minimum to maximize reliability and to minimize thermal and structural perturbations and cost. Five unavoidable mechanisms are the CCD shutter, the modulator drive to rotate the modulator to 12 angular positions over a full rotation, a PSP channel select fail-safe mechanism, an aperture door frangibolt and hinge, and the secondary mirror tip/tilt and focus mechanism.

A calibration source module comprising of Pt-Ne hollow cathode lamp, a Deuterium lamp and a Krypton lamp are in the current baseline to enable spectral and flat field calibration.

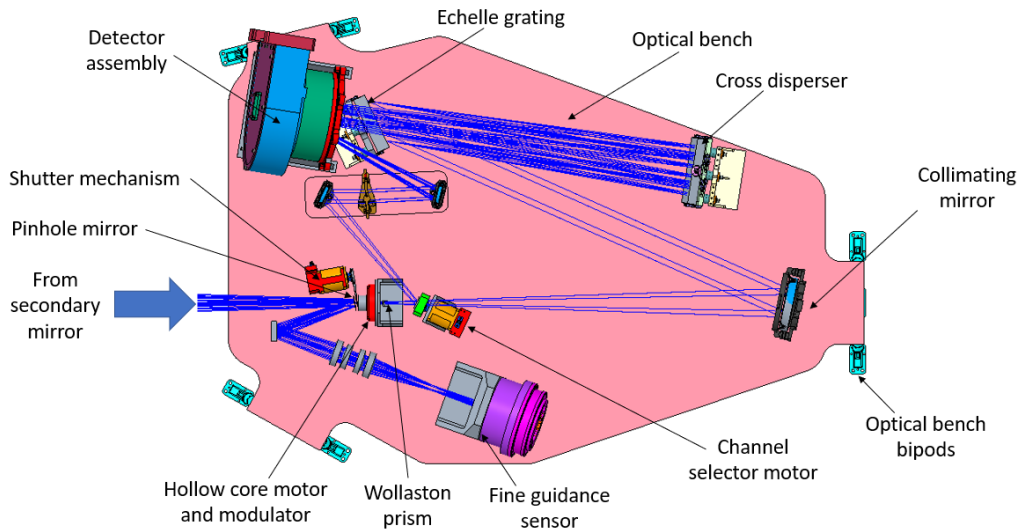

Figure 2 - An inline spectropolarimeter layout enables better control of the system errors and simplifies assembly, integration and test.

## 2.2 Polstar Instrument Functional Approach: the Polarimeter

The heart of the instrument design is the rotating modulator that allows isolation of individual polarization planes from the input light from the OTA. The design uses a stack of $MgF_2$ crystals to take advantage of the birefringence of the material to isolate the entrance polarized wave and pass it through the system. The design then uses a second analyzer crystal at fixed orientation, a Wollaston prism, to split the entering wave into two beams and encode the polarization into the intensity of the signal passed by the analyzer, Figure 3.

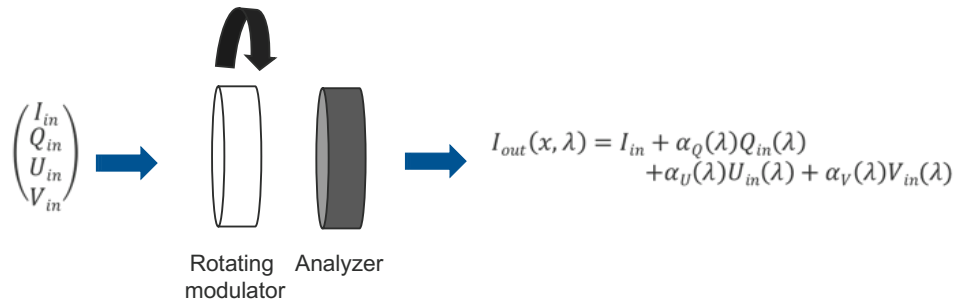

$$\begin{pmatrix} I_{in} \\ Q_{in} \\ U_{in} \\ V_{in} \end{pmatrix} \rightarrow \text{Rotating modulator} \quad \text{Analyzer} \rightarrow \begin{aligned} I_{out}(x, \lambda) &= I_{in} + \alpha_Q(\lambda) Q_{in}(\lambda) \\ &+ \alpha_U(\lambda) U_{in}(\lambda) + \alpha_V(\lambda) V_{in}(\lambda) \end{aligned}$$

Figure 3 – Pictorial representation of the principles of intensity encoding the input polarization as represented by the 4 Stokes vectors of the incoming beam.

The idea then is to use the multiple positions of the rotating modulator to provide the full characterization of the input beam, as shown in Figure 4. The parameters of the plates (thickness, orientation) and modulator angles are optimized to obtain the best efficiency in the extraction of the polarization information.

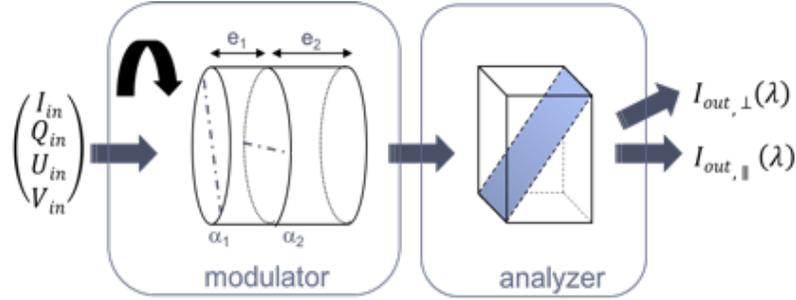

Figure 4 – the Polstar implementation of the principle uses multiple thin MgF2 crystals to build the rotating modulator and a second MgF2 crystal with a specific tilt angle on the birefringent axis to separate two beams that allow a differential measurement at the focal plane that removes systematic errors in the system.

## 2.3 Design Introduction

The dual-channel Polstar MIDEX design will measure the Stokes 4-vector $(I, Q, U, V)^T$ of selected stars in the FUV Channel 1 ($\lambda\lambda$ 122 nm to 200 nm) and the NUV Channel 2 ($\lambda\lambda$ 180 nm to 320 nm) with low and high spectral resolving power, respectively, [> 30K for Channel 1 and > 30 for Channel 2]. This section will describe the optical system trades to define entering into the design of this high-precision, large spectral resolving power, space-borne, cost constrained spectropolarimeter.

**Table 1** outlines the top-level design requirements: desired and derived. To measure polarization state and degree of polarization (DoP) with high precision requires design features and constraints that do not apply to non-polarimetric systems. The design elements provide light collection, spatial isolation, polarization sensing, spectroscopy, and read-out. The elements that provide these functions (see Figure 5) are the Optical Telescope Assembly (OTA), Entrance Slit (ES), Polarization Assembly (PolA), spectrometers, and Focal Plane Array (FPA), respectively. Taken in this order, the combined ES, PolA, spectrometers, and FPA are grouped into the Polstar Spectropolarimeter (PSP).

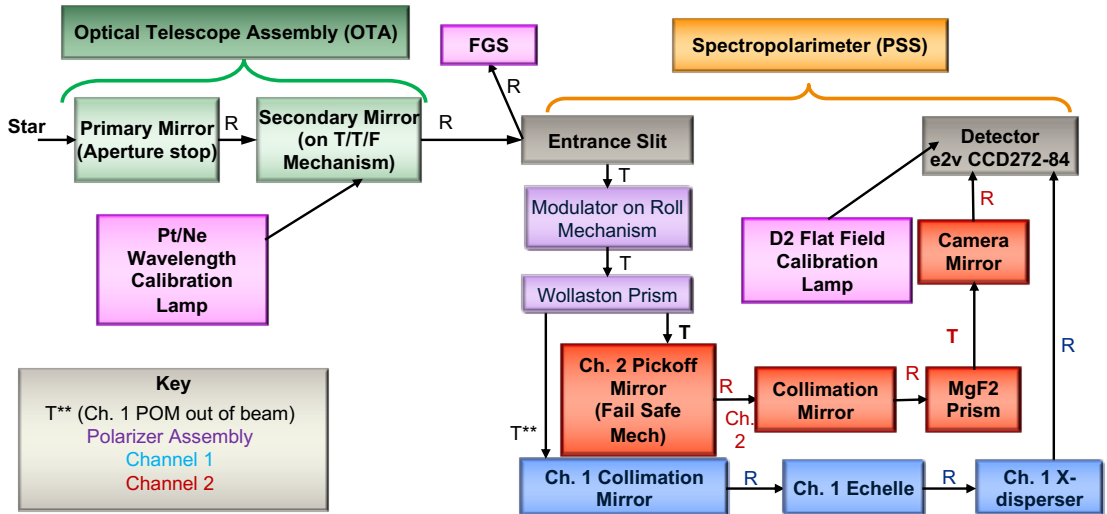

Figure 5 – Optical Block Diagram design of the Polstar Instrument.

Precise polarimetry requires that any optic that the incident light encounters before the PolA does not modify the stellar sources polarization state and DoP. If the Stokes vector $\mathbf{S_s} = (1, q, u, v)^T$ represents the polarization properties of the stellar source, the design must strive to present $\mathbf{S_s}$ as accurately as possible to the PolA. Any property of the OTA which could modify $\mathbf{S_s}$ must be controlled in the basic design and in its implementation. The Polstar design implements this by full axial symmetry, minimized number of mirrors, low angles of incidence, no fold mirrors, control of reflective coating symmetry, and full symmetry of obscurations. Efficiency in the FUV spectral range also requires minimizing the number of mirrors. In addition, the cost constraints of a MIDEX mission puts limits on OTA aperture size.

Referring to **Table 1**, Polstar uses a 60-cm diameter f/13 True Cassegrain OTA with relatively slow f/2 PM and sufficient back focal length (400 mm) for PSP packaging. The f/13 beam focuses the field (star) on the shared ES. The PSP incorporates two separate spectropolarimeter designs: Channel 1 for R > 30 K over the FUV 122 to 200 nm waveband and Channel 2 for R > 30 over the NUV 180 to 320 nm waveband. [As implemented Channel 2 is usable at wavelengths from the $MgF_2$ blue cut-off (115nm) to the CCD red cut-off (1060nm) with R dependent on $MgF_2$ dispersion with wavelength.]

Channel 1 uses a Hubble Space Telescope spectrometer design echelle spectrometer with optimized grating cross-disperser to achieve the R > 30K spectral resolving power. The R of Channel 2 is readily achievable with a prism spectrometer again with HST heritage. MIDEX cost constraints dictate minimizing the number of detectors, so the designs share the same single science FPA. The design implements this with a single mode select mechanism that inserts a flat mirror in the common path to direct the light to the Channel 2 collimator, prism, and camera mirror. The two channels are thereby used separately as the mode select mirror blocks the beam to the Channel 1 spectrometer optics. The two channels share the OTA, the entrance slit, the PolA, and the FPA.

Of prime importance is the PolA. Our design uses heritage designs from the experience of co-authors Neiner and Casini. The design consists of two key elements each of which provides a distinct function. The first PolA element is the modulator with two $MgF_2$ retarders (pair 1 and pair 2), each of which consists of two retarders, i.e., each is constructed as a pair of retarders. The optic axis (fast axis) of each of the four retarders are normal to the optical axis (Polstar OA) of the incident light. The individual optic axis of the individual retarders in each pair lies at 90 degrees to each other. This creates a net retardance of the pair solely dependent on the physical thickness *difference* between the individual retarders (and wavelength as well as birefringence of the retarded material). The two retarder pairs are mounted with a fixed azimuth angle, roll angle about the Polstar optical axis, relative to each other. The second PolA element is a $MgF_2$ Wollaston prism analyzer.

In operation, the modulator retarders rotate about the Polstar optical axis, but the Wollaston analyzer is fixed. A single observation consists of a set of single integrations at each of six retarder roll angles. The resultant encoding of the polarization signal yields six measurables that we use to solve for four unknown $(I, Q, U, V)^T$. Each modulator output is polarized into s and p polarization states and slightly diverted angularly, in the cross-dispersion direction, by the Wollaston prism. The resultant beam, dispersed by the spectrometer, forms two full spectra on the FPA displaced relative to each other.

## 2.4 Optical system trades and configuration

The overall configuration requires a flux collecting telescope that feeds incident light into spectrometers which in-turn provide the requisite spectral resolving power (R) when sampled by a suitable focal plane array (FPA). MIDEX mission cost constraints set the aperture size to 60-cm diameter or smaller. Polstar adopted an aperture diameter, D, of 60-cm. Efficiency across the FUV to NUV spectral region requires an all-reflective Optical Telescope Assembly (OTA), not a refractive OTA. The UV wavelengths require minimizing the number of reflections and slow f/# limiting the choice to two-mirror OTA designs. The chosen two-mirror Cassegrain design form minimizes overall length, compared to Gregorian designs.

The Cassegrain OTA feeds suitable spectrometer designs with the required polarization modulator and analyzer placed after the entrance slit. Any such polarizing elements, the Polarization Assembly (PolA), will be small physically compared with the OTA size. They need to be near Cassegrain focus to minimize their size and provide the required amount of separation of the s and p polarization spectra at the FPA.

The basic design and its implementation eliminate the sources of polarization measurement errors above the allowed 0.1% per exposure polarization precision and accuracy and calibrates residual errors remaining in the OTA and the spectrometer. Higher precision can be achieved by stacking derived Stokes products for the same target after the fact. For example, any optic or vignetting structure forward of the PolA can introduce uncompensated polarization to the incoming stellar signal, unless they possess precise axial symmetry. Axial symmetry of angle of incidence, coating reflective efficiency and phase, figure mid-spatial frequencies, and blockages effectively cancels polarization from the opposite region of the aperture. Therefore, the OTA must be used on-axis in aperture and on-axis in field, and therefore be centrally obscured. Its coatings and obscuration must have exquisite axial symmetry. Therefore, the secondary mirror must be supported by four straight spiders (or zero).

Table 1 Polstar Top Level Requirements

| Item | Requirement | Design | Notes |
|---|---|---|---|
| **OVERALL:** Two-mirror Cassegrain telescope feeds two spectropolarimeters that share Entrance Slit, CCD shutter, Polarization Assembly, and CCD | | | |
| OTA design form | Small FOV; High UV efficiency | True Cassegrain | PM: CC parabola, SM: CX hyperbola |
| OTA implementation | Axial symmetry wrt optical axis | Complete symmetry to null induced-pol. | Centrally obscured, 4-arm spider; Fully baffled (Baffles: SM cone, PM central, Outer barrel) |
| OTA aperture D | ≤ 600 mm | 600 mm | Achieve science: In MIDEX cost limits |
| OTA f/#, PM f/#, BFL | ≥ f/13 | f/13; f/2; BFL = 400 mm | Polarization Assembly heritage (f/13); Reduce OTA induced-polarization (f/2); Permit stable strongback design (BFL) |
| Sun avoidance angle | ≥ 95 degrees | 95 degrees | View nearly $2\pi$ steradians: 95 degrees of solar vector to anti-sun |
| Spectrometer | 2 channels | 2 channels | Channel 1 (FUV), Channel 2 (NUV) |
| Ch. 1 and Ch. 2 share | | Share OTA, ES, shutter, PolA, FPA | Minimize cost and complexity. Achieve science capability |
| Entrance slit | For spectral purity | 793 mas dia. | Two channels share; Largest allowed by spectral resolving power |
| Channel 1 (λλ nm) | λλ 122 to 200 | Same | All reflective OTA/spectrometer designs |
| Ch. 1 R | R ≥ 30,000 | 32.4 K to 34.3 K | 2-pixel sampling. Orders 37 to 60. |
| Ch. 1 PSP design | | X-dispersed echelle | Heritage: HST/GHRS, HST/STIS. X-Disperser in Wadsworth configuration |
| Channel 2 (λλ nm) | λλ 180 to 320 | Same | |
| Ch. 2 R | R ≥ 30 | R > 34 | 2.5-pixel sampling |
| Ch. 2 PSP design | | Prism spectrometer; MgF$_2$ prism | All reflective except prism |
| Science FPA | | e2v, CCD272-84 | Two channels share; 4K x 4K, 12-micron pixels |
| **POLARIZATION ASSEMBLY & MEASUREMENT:** Each of 4 Stokes parameters: Linear and circular polarization of stellar incident light. (I, Q, U, V)$^T$ | | | |
| Polarization Assembly (PolA) | Measure all 4-Stokes parameters | Roll retarder about Polstar optical axis; Fixed Wollaston prism analyzer | Science team heritage. Six modulator roll positions encode incidence polarization; Wollaston polarizes s and p creates dual spectra simultaneously on FPA |
| Polarization Accuracy/Precision | < 0.001 | Design to 0.001 | Polarization measurement accuracy and precision of system < 1E-05 using data binning. |
| Polarization calibration | Full characterization on ground | Ground (1,0,0,0)$^T$ spectral source: OTA and PSP cal. | Special source: calibrate pol. degree and state, wavelength, spectral resolving power, SNR. Use stellar sources in-flight |
| Image shift at FPA | ≤ 1-micron | PSP optical bench alignment stable | Over one hour. PSP thermally & mechanically isolated. Drives mechanical/thermal design. Kinematically attach PSP to strongback. |
| **FINE GUIDANCE SYSTEM & SENSOR** | | | |
| FGS FOV | ≥ 8 arc min dia. | 8 arc min dia. | |
| FGS view region around science star | | Pierced mirror | Passes light to PSP ES. Reflects FOV to FGS. |
| FGS detector | | 2K x 2K CIS CMOS | 10-micron pixels, f/14.7. λλ 500 to 1000 nm; 234 mas pixel on sky |

The f/# of each mirror must be slow enough to introduce little polarization sensitivity to the incoming beam. We choose f/2 for the PM based on these trades and set the OTA f/# at f/13 consistent with experience with the selected PolA components. The Aperture Stop is at the PM. We also limit misalignment errors of the PM and the SM to negate introduction of instrumental polarization. We choose to place the PolA behind the entrance slit inside the spectrometer, not before the entrance slit, to protect it from stray light and from temperature extremes if it views cold space. Thus, the image of the field at Cassegrain focus encounters no refractive elements and is free of chromatic aberration.

Achieving spectral resolving power in the R ~ 30 K range in Channel 1 requires an echelle spectrometer form. Based on experience from HST/GHRS and HST/STIS, the all-reflective design collimates the diverging beam from the entrance slit reflecting it to the plano echelle grating used at sufficient angle of incidence to achieve R > 30K. An imaging cross-disperser intercepts the diffracted beam from the echelle dispersing each order separate from its adjacent order and images the spectra as a classical echellogram. The design uses orders m = 37 through 60, so the echellogram consists of 24 separate orders (per polarization state). The cross-disperser is used in an aplanatic "Wadsworth" configuration, with the mid-wavelength of the overall spectra near zero angle of diffraction, to hold third-order coma to zero at the mid-wavelength, and to minimize coma over the full echellogram footprint.

The low-resolution Channel 2 spectrometer uses a prism disperser, a $MgF_2$ prism, with planar surfaces, used at minimum deviation. A collimation mirror collimates the beam to the prism and a camera mirror images it on the detector. The two channels share the one detector to minimize cost. The channels do not operate simultaneously. A mechanized flat pick-off mirror intercepts the beam to Channel 1 directing beam to Channel 2.

Key to the measurement of the source state of polarization, $S = (I, Q, U, V)^T$, is the design and implementation of the Polarization Assembly. It must precisely measure FUV and NUV 4-vector parameters in a compact design suitable for space-mission deployment. [13] outlines the adopted general approach that uses a rotating retarder of phase difference, $\Delta$, followed by a fixed analyzer. The rotating retarder encodes the incident state of polarization onto the exiting flux modulating all four Stokes parameters, including circular polarization. With the state encoded, polarization properties of optics following the PolA do affect the spectral throughput, but not the polarization measurement.

Neiner has refined this approach [10] by using a retarder with two waveplates with optimized retardance. The optical axes of the crystals are normal to incident flux and rotated to each other by a fixed angle. Consistent with FUV and NUV wavelengths, the retarders are single crystal $MgF_2$. The retarder-pair rotates 360 degrees about the system optical axis stopping at six selected azimuth angles per 180-degree rotation. An exposure of selected temporal duration is recorded at each selected azimuth angle. A single observation consists of the six set of exposures.

The approach uses a fixed $MgF_2$ Wollaston prism to analyze the output of the rotating retarders yielding simultaneous sampling of the modulated signal in separate orthogonal polarization states. The output angle between the two states (s, and p polarization) at each wavelength is fixed and set by the spectrally dependent birefringence properties of the Wollaston prism material and the prism angle of the two halves of the prism. This angular deviation yields a virtual displacement forming two virtual entrance slits: one for each polarization from the Wollaston. The spectrometer then images these as two individual spectra at the FPA resulting in the simultaneous formation of two displaced spectra on the spectropolarimeter focal plane. By design, the Wollaston deviation is in the cross-dispersion direction. The resulting Channel 1 echellogram becomes two interleaved echellograms with the polarization displaced images occupying the region between the nearest spectral order. The Channel 2 spectra becomes two parallel spectra of orthogonal polarization. Note that if the source intensity were to vary, the signal of both spectra varies proportionally. By ratioing the two measurements the determined polarization states are independent of source fluctuations.

Each observation consists of integrating the pixel output for each of the six exposures. Image location over this time period, up to one hour of signal integration, must be maintained to ~ 1.2-micron, 10 % of a pixel, to ensure intra-pixel sensitivity variations do not introduce cross-talk among the Stokes parameters. The design as implemented will ensure such stability. The spectropolarimeter (PSP) optical alignment must be maintained, even at the 20-mas level of optic tilt. Thermal and mechanical isolation achieves this by mounting the spectropolarimeter kinematically to the OTA and only to the OTA. The OTA is in turn mounted kinematically to the spacecraft. These isolation measures ensure that spacecraft and OTA-induced flexures do not flex the optical alignment within the PSP. Any such perturbations will move the PSP as a solid unit with no internal flexing. The PSP is thermally isolated radiatively and conductively from the environment to ensure stability of temperature gradients over the 1-hour period.

Line-of-sight (LOS) jitter and drift tilts the OTA and thereby translates the image of the field at Cassegrain focus. The PSP entrance slit is a pinhole 30-micron in diameter, 800-mas in field, placed at Cassegrain focus. This dimension maps into 2-pixels, 24-micron in Channel 1, and 2.5 pixels, 30 microns in Channel 2. The spectral resolving power is set by these pixel-sampling values. Therefore, a field angle LOS offset > 40-mas, if relayed to the detector, would shift the spectral image by full allowed tolerance of 1.2-micron. If the system design imposed an error of this magnitude, we would not be allowed to image the star directly at the entrance slit. We would need to defocus the OTA and overfill the entrance slit so LOS motion would not shift the image at the detector. However, LOS jitter will average to zero over the integration period, so it will not shift the recorded image centroid. Lower temporal period LOS drift would not average to zero and would introduce a net offset.

In addition, LOS motion is only one of a myriad of error sources as described above. We determined the allowed contribution of each possible error on the overall image shift with an Image Stability Error Budget that includes each error source and their contribution to the image shift. The error budget then combines these statistically. LOS jitter and drift are terms in the error budget. The error budget confirms that with LOS drift < 20-mas, we can image the field sharply at the entrance slit and still maintain image motion at the FPA < 1.2-microns. The Pointing Control System is designed to provide a zero-average jitter and LOS drift < 20 mas, values that when combined with other errors meets the 1.2-micron requirement.

## 2.5 Calibration trades

Polstar will require a comprehensive calibration both pre-flight and during flight. These include wavelength calibration, photometric calibration, and polarimetric calibration. The first two are standard for such spectrometric instruments. In-flight an on-board Pt/Ne hollow cathode wavelength calibration source will calibrate the pixel locations on the FPA of known wavelengths. Standard stars and ISM lines will also provide calibration in-flight.

Unique to Polstar is the rigorous pre-flight ground polarimetric calibration. The PSP will be calibrated viewing known 4-vector polarization state inputs generated by a ground calibration source. Full recovery of Stokes vectors will be demonstrated with known input of linear and circular polarization across the spectral region. In addition, polarization sensitivity of the OTA will be characterized.

## 2.6 Returning FUV Spectropolarimetry after 30 years absence

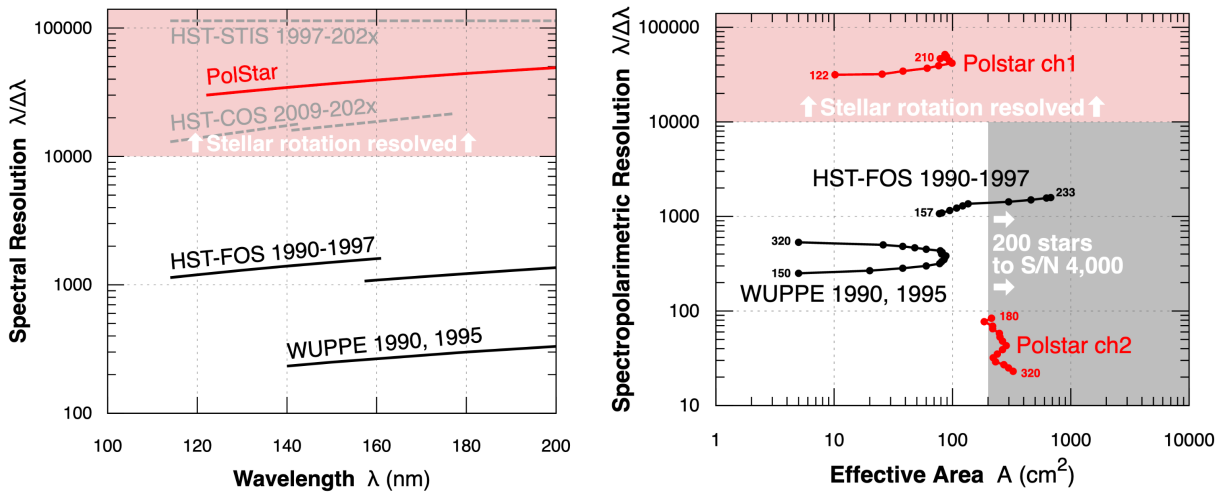

Figure 6 – A comparison of the projected performance of the Polstar instruments with their predecessors on the Hubble Space Telescope, and the WUPPE mission. Comparable levels of spectral resolution are delivered across the UV band while at the same time delivering orders of magnitude improvement in parallel polarimetric capability at comparable effective areas, despite the more modest aperture size, leveraging advances in new technology and design.

Polstar uses all the design and technology innovations discussed above to deliver a new capability in the FUV – the combination of high resolution spectroscopy combined with high precision polarimetry that takes advantage of the numerous diagnostic lines available in the FUV band. The mission as proposed will return FUV spectropolarimetry to the astronomical community for the first time in 25 years to deliver (Figure 6) comparable spectral resolution and effective areas to previous HST instruments, but add the vital new capability of time domain polarimetry at a resolution orders of magnitude higher than has ever been possible before, and at high precision. The advances this mission will enable will be transformative.

## 3. THE BASIC SCIENCE OF POLSTAR

Understanding how a galaxy such as ours can evolve to support life requires understanding stellar feedback into star formation and the interstellar medium, and the processes that produce planets and promote their habitability. Though rare, massive stars are important because their impact is felt violently and immediately in the star-forming environment, whereas lower-mass stars provide a steadier and more global influence in the galaxy as a whole. The latter also incubate life by forming and maintaining planets. Our primary means of understanding these processes is by analyzing starlight that varies in intensity as a function of wavelength, angle (in the rare cases when spatial imaging is possible), polarization, and time.

A great deal of astronomical technology has already been aimed at the ultraviolet wavelength regime, and less often, angular resolution, but the domains of polarization and time variability remain largely unexplored. Though often at only a percent level or less, the degrees of polarization and time variability provide uniquely powerful constraints about the geometry and physical processes at play, as described further here. The purpose of Polstar, then, is to access this unique information by combining high spectral resolution (resolution at the 10 km/s level) and high precision (0.01% scale) polarimetry with the ability to delve deeply into the time domain on a smaller number of key benchmark targets, to complement the extensive existing archive of spectral snapshots of a wide array of targets that largely leave behind this time-domain and polarization information.

This paucity of detailed constraints on geometry and dynamical variations is especially pronounced for massive stars, for whom our extensive familiarity with the Sun is less relevant, and could therefore exhibit types of phenomena about which we remain as ignorant as we once were of Galileo's sunspots. Massive stars are generally distant point sources, so spatial resolution must be replaced by proxies such as frequency dependence induced by Doppler shifts in flows, and polarization induced by aspherical geometry. Also, the high luminosity of massive stars produce radiatively driven winds that remove so much mass it can alter their evolution and their impact on their galactic environment. In addition to steady winds, massive stars often partake in more impulsive events such binary mass transfer and eruptive incidents, which also alter their fates and their roles in a galaxy. Understanding the physics of these flows, and quantifying their impact, requires studying these variable behaviors on timescales from months to hours, and we cannot understand their physics, or properly constrain their mass fluxes, without understanding this behavior and its mysterious causes.

Polstar will approach these questions using dynamical spectra that are well resolved in wavelength and time, and augmented with polarimetric constraints, to probe wind structure and correct the mass-loss rate for the effects of wind clumping and asphericity. Also, the processes that spin up and spin down these stars will also be studied from a geometric and time-dependent perspective, as angular momentum evolution also affects the nature of the endpoints and supernova types, such as gamma-ray bursts and black hole binary mergers. The role of magnetism will also be considered by measuring circular polarization from the Zeeman effect, where knowledge of the photospheric magnetic field and the modification of the circular polarization induced by the hypersonic acceleration of a wind uniquely constrains models of wind initiation and informs the causes and properties of massive star magnetism. Let us present in some further detail how the attributes of Polstar allow certain key questions involving massive star evolution to be addressed, before going on to the questions involving feedback into the interstellar medium, and the properties of planet formation.

### 3.1 Polstar massive-star science

Polstar will address massive-star science in regard to five key objectives. The first objective is to understand aspherical and time varying structures in the wind that alter our determinations of the stellar mass-loss rate enough to change our predictions for the star's evolutionary pathway. Small-scale structure can be probed via dynamical spectra by staring at the star for up to a week, and watch the evolution of wind structures in real time. Past HST and IUE observations [14] show that such time variability is ubiquitous in massive-star winds, for mysterious reasons, and it is also known that such winds exhibit substantial small-scale clumping (and generate X-ray hot shocks), but the effects of these structures on our inference of the mass flux can only be estimated without a better understanding of their causes and behaviors. Polstar will track wind velocity patterns as they accelerate through the line profile, achieving for a canonically bright target with intensity $10^{-10}$ erg s$^{-1}$ cm$^{-2}$ Å$^{-1}$ a signal-to-noise ratio of 300, over spectral bins of 30 km/s and for exposure times of about twenty minutes, appropriate for the timescale for the features to accelerate across the spectral bins in question. For a target 4 times brighter, the spectral bin and exposure time could both be halved at the same signal strength, providing even higher fidelity, and dozens of such sources exist. Also, linear polarization from longer exposures and the much wider frequency bins of channel 2 can be used to probe more global and time steady wind asphericities, such as could be produced by rapid rotation or magnetism, providing independent constraints on the wind mass flux than what is available spectroscopically.

The second objective is to track the processes that spin up massive stars, as many are rapid rotators and some even rotate so fast that their equatorial surface region is nearly in orbit and given to creating equatorial disks [15,16]. Binary mass transfer is hypothesized to be the crucial process that can achieve this despite the constant angular momentum sink coming from the stellar wind [17], but many rapid and critical rotators are not yet known to be in binaries. Here Polstar can look for unseen companions by looking for spectral features and linear polarized position angles that recur on a strict period, attributable to an orbiting companion. Again it is Polstar's ability to obtain spectropolarimetric time-domain constraints that will be decisive, the polarimetry again coming from channel 2 where wide wavelength bins will be used to obtain the necessary photon counts. Polstar's effective area can achieve a signal-to-noise of 20,000 with hour-long exposures on canonically bright ($10^{-10}$ erg s$^{-1}$ cm$^{-2}$ Å$^{-1}$) massive stars with a channel 2 resolving power of 30, using the wavelength dependence and searched-for binary periodicity to separate the stellar from interstellar polarization.

The third objective also involves binarity, but this time the objective is to understand the degree to which mass transfer is conservative, and how that affects the stellar endpoints in the system. Here Polstar can look for spectroscopic evidence of stellar rotation that is unsynchronized, implying that mass transfer is occurring too quickly to maintain tidal locking and is overloading the gainer with angular momentum [18,19]. Phase-locked polarization in channel 2 would look for evidence of escaping mass, which would be highly aspherical. Mass flows seen in projection against the star would also produce spectroscopic evidence in the high-opacity UV resonance lines, which could be detected at a signal-to-noise of 300 in 10 km/s bins for canonically bright sources seen in hour-long exposures with phase coverage over the orbital period of days to weeks, exposing the nature of binary mass transfer.

The fourth objective is another binary-related objective where the presence of an orbiting companion provides a unique probe of a hot-star wind by providing a source of light and mass flux seen from varying lines of sight through it. Here the purpose is to understand the radial dependence of the density and velocity in the wind, and again a signal-to-noise of 300 in 10 km/s bins is achieved similarly to the previous case. Channel 2 can again be used to detect phase-locked polarization owing to the changing geometric perspective. It is apparent that significant overlaps in the experiments described will sometimes allow multiple objectives to be carried out on the same targets, with the same data, reducing required mission time.

The final massive-star objective is to study previously observed strong stellar magnetic fields to establish their effect on wind dynamics, both by restricting mass loss by an order of magnitude, and inducing rapid spindown of the star [20,21,22]. Here the crucial goal is observing the Zeeman effect in circular polarization, induced low in the wind where the magnetic fields are still enough and the Doppler shifts weak enough that the ratio of Zeeman shift to Doppler shift is not too small to allow us to detect the correlations between them. Those correlations produce circularly polarized structure in the line profile that deviates from the standard result (proportional to the wavelength derivative of the line profile), providing us with unique and independent information about the early acceleration of the wind. For this correlation to be detected by Polstar the magnetic fields must be in the vicinity of 10 kilogauss, and the instrument must co-add wavelength domains from a number of different lines that respond with similar physics. Using an effective area of 40 cm$^2$ in channel 1, and requiring a signal-to-noise ratio of 2000 for a 10 kilogauss field to be detectable, we find that data bins of roughly 400 km/s are required for Polstar to collect enough photons per bin. Since the features in each line should require about 50 km/s resolution to characterize, this implies ~8 reasonably strong resonance lines that involve similar line-formation and Zeeman-shift physics will need to be co-added, but many such lines are present in the UV. Since the field needs to be quite strong, Polstar will only be able to address a few sources and may be considered a kind of proof of concept of this approach to understanding the magnetic field environment and the initiation of wind acceleration. The ability for Polstar to measure linear polarization also offers the potential to use the Hanle effect to probe much weaker fields on the scale of 100 Gauss, as is often done in observations of the Sun, but this type of modeling [23] is in its infancy when applied in the spatially unresolved stellar context.

## 3.2 Polstar interstellar medium science

Ultraviolet (UV) spectropolarimetry (primarily in channel 2) also provides new constraints on the smallest dust grains in the interstellar medium (ISM), which play an important role in galactic chemistry, star formation, and the propagation of ionizing radiation. Here Polstar's mission will center on five objectives, testing how FUV and EUV align the smallest grains in a magnetic field, how different EUV environments can lead to a difference in the grain size at maximum alignment, how this depends on the metallicity in the chosen sightline, and what is the source of the mysterious 2175 Angstrom extinction feature. The fifth objective relates to ground state alignment in interstellar atoms due to the orientation of the magnetic field relative to the angle toward the illuminating source. Only this final objective involves atomic resonances and the high spectral resolution of channel 1.

These objectives involve how ISM dust and gas respond to magnetic fields and ionizing radiation. UV spectropolarimetry allows probes of the smallest dust, which has the most significant impact on the short wavelengths of ionizing radiation. This connects with radiative reprocessing, understanding cosmological backgrounds, and constraining the degree of ionization in the ISM. EUV radiation shortward of the Lyman limit plays an important role, is difficult to observe directly, and along with magnetic fields, plays a key role in regulating galactic plasma dynamics. The containment and escape of galactic plasma echoes the containment and escape of magnetized stellar winds, so Polstar will simultaneously inform these processes on widely different scales.

All of these goals relate to UV ISM polarization levels up to a few percent, and thus the sensitivity requirements are less stringent than for the massive-star objectives. Nevertheless, observations of both types are involved for each objective, because of the need to separate and understand stellar and ISM polarization sources to be able to understand either. The different wavelength dependences of dust and atoms, the narrow width of ISM lines, and the generally time steady nature of ISM polarization, makes the separation straightforward. Again it permits different objectives to be addressed by the same observations.

## 3.3 Polstar exoplanet science

The final area of emphasis of Polstar are the processes that assemble and maintain exoplanets, and contribute to transient events that could challenge prolonged habitability. One type of transient event, atmospheric blowoff events triggered by stellar activity, will be explored entirely spectroscopically, and requires a relatively low signal-to-noise of around 10 in order to correlate the appearance of absorption lines during ingress or egress of transiting planets with strong stellar flares in active dwarfs. Owing to the low signal level required to distinguish transient line events, blowoff events at 10 km/s resolution can be detected transiting the brightest active dwarfs in impulsive minute-long timescales, whereas dimmer targets requiring longer exposures will only be selected if they exhibit more extended periods of strong flares conducive to longer exposures. These diagnostics extend from the UV to the optical, a dynamical range showing a substantial range in stellar brightness that allows the red end of channel 2 to access the highest signal-to-noise. To fully utilize the red end of Polstar's spectral range may require strategies aimed at avoiding saturation of Polstar's CCD detector, such as spreading out the signal over more pixels than in the current channel 2 design.

Polarization in channel 2 will be utilized for the two objectives relating to planet-forming disks. The first emphasizes the polarization diagnostics of the theoretically expected break from magneto-accretion to boundary-layer accretion, to quantify the host mass where this break occurs [24,25,26]. The second looks at transient events in the inner regions of the disk, too close to the star to be spatially resolved, to distinguish inner-disk misalignments that herald planet formation from evaporation events that herald planetary destruction [27,28]. This latter set of experiments couples channel 2 polarimetry of the geometric transients with channel 1 spectroscopy in narrow spectral lines accessible at Polstar's 10 km/s resolution. Both these objectives can be achieved in exposures of a few minutes for brighter targets, as the expected polarization transients are at the percent level or just below, so can be obtained via the wide spectral bins of channel 2, while the spectroscopic signals require only enough signal-to-noise to correlate their appearance and disappearance with orbital periodicities in the protoplanetary disk. A fourth objective, involving probes of habitability and chemistry based on reflected light from the planetary atmosphere, is under consideration, but requires polarimetric precision at the 0.001% level, and may prove too technically challenging.

## 4. CONCLUSIONS

The Polstar mission design and science program represent the convergence of new advances achieved via technological investment development in the ultraviolet, with advances in polarimeter design, and scientific advances in the knowledge and understanding of the environments of massive stars, the interstellar medium and the formation of exoplanetary systems. Polstar will return the capability of FUV spectropolarimetry to the astronomical community after a 25 year absence, and do it with orders of magnitude improvement in the combination of sensitivity and spectral resolution, while delivering unprecedented polarimetric accuracy in the diagnostic-rich domain of the ultraviolet. The mission is responsive to priorities identified in the 2010 Decadal Survey that outlined the need for a better understanding of the influence of stellar environment and structure on the ultimate evolution of the single class of star that most influences the baryonic cycle that shapes the stellar, planetary and chemical evolution of galaxies. We look forward to making our proposal to NASA at the end of 2021.